\documentclass[%preprint,           % Format : preprint, twocolumn
               twocolumn,
               noshowpacs,          % Pacs : showpacs, noshowpacs
               nopreprintnumbers,     % Preprint: preprintnumbers,
               aps,                 % Society: 
               prd,                 % Journal Style : pra, prb, prc, prd, pre,
               a4paper,             % Size : a4paper
               superscriptaddress,  % Affiliation (Title) : groupedaddress,
               nofootinbib,         % Footnote: footinbib, nofootinbib
               tightenlines,        % Remove additional spaces in a line
               floats,floatfix      % Floating pictures and tables
               ]{revtex4}

\usepackage{amsmath, amsfonts, amsthm, amssymb, graphicx}
\usepackage{epstopdf}
 \usepackage{slashed}
 \usepackage{graphicx,epsfig}
\usepackage{amsmath}
\usepackage {amssymb}
\usepackage[utf8]{inputenc}
\usepackage{hyperref}
\usepackage[francais]{babel}
\usepackage[T1]{fontenc}
\newcommand{\be}{\begin{eqnarray}}
\newcommand{\ee}{\end{eqnarray}}
\newcommand{\bea}{\begin{eqnarray}}
\newcommand{\eea}{\end{eqnarray}}

\begin{document}

\title{Higher-curvature corrections and near horizon symmetries}

\author{Mariano Chernicoff}
%\email{}
\affiliation{Departamento de F\'{\i}sica, Facultad de Ciencias,
Universidad Nacional Aut\'onoma de M\'exico,
A.P. 70-542, CDMX 04510, M\'exico.}

\author{Gaston Giribet}
%\email{}
\affiliation{Department of Physics, New York University, 726 Broadway, New York, NY10003, USA.}

\author{Julio Oliva}
%\email{}
\affiliation{Departamento de F\'{\i}sica, Universidad de Concepci\'on, Casilla, 160-C, Concepci\'on, Chile.}

%\pacs{}

%\date{\today}

\begin{abstract}
In the near-horizon region, black holes exhibit an infinite-dimensional symmetry reminiscent of the Bondi-Metzner-Sachs (BMS) supertranslations. The conserved charges associated with this symmetry can be computed in gravitational theories of arbitrary spacetime dimension and involving curvature terms of any order. In Lovelock theory, for instance, these charges take the form of nested Lagrangian densities corresponding to topological invariants, each weighted by the supertranslation function -- thus providing a natural generalization of the Wald entropy formula. In four dimensions, the computation of the supertranslation charge reduces to the evaluation of the Jackiw-Teitelboim (JT) action on the two-dimensional spacelike sections of the event horizon.

\end{abstract}

\maketitle

\section{Introduction} 
\label{introduction}

Almost a decade ago, Hawking proposed that black holes might exhibit infinite-dimensional symmetries near their event horizons \cite{Hawking}. More specifically, he suggested that, much like the asymptotic region near future null infinity in asymptotically Minkowski spacetimes, the geometry close to a black hole event horizon should display a supertranslation symmetry akin to the BMS symmetry. This idea, which had earlier precedents \cite{Hotta, Koga, Hotta2}, was later formalized by Hawking, Perry, and Strominger in \cite{Strominger}.

In \cite{Donnay}, the authors demonstrated explicitly and in a general setting that black holes indeed exhibit not only supertranslation symmetry near the horizon but also superrotation symmetry. Moreover, the associated Noether charges carry significant physical information about the gravitational solution \cite{Strominger2, Donnay2}; see also \cite{Grumiller} and references thereof. For instance, the zero mode of the supertranslational charge corresponds to the Wald entropy \cite{Wald}, while additional conserved charges —computable directly from the horizon— display a kind of Gauss phenomenon.

This framework has since been applied across various contexts, spacetime dimensions, and setups. However, with a few exceptions \cite{Pu, Donnay3}, most analyses of near-horizon symmetries have been limited to Einstein gravity. The goal of this brief note is to highlight that a similar symmetry analysis is possible in more general theories of gravity, including those with higher-curvature corrections. To that end, we consider Lovelock theory, a well-known example that extends general relativity by incorporating higher-order curvature terms in a way that yields second-order field equations. We show that, in arbitrary spacetime dimension and for any number of curvature terms, the supertranslation charge at the horizon can be computed explicitly. These charges take the form of nested Lagrangian densities corresponding to dimensionally continued topological invariants, each weighted by the supertranslation function -- thus providing a natural generalization of the Wald entropy formula. In four dimensions, provided the Euler characteristic is kept in the action, the computation of the supertranslation charge reduces to the evaluation of the JT action on the two-dimensional spacelike sections of the event horizon. A similar result holds for both extremal and non-extremal horizons. In the former case, however, the infinite-dimensional symmetries are interpreted not as supertranslations but as superdilations acting on the null-time coordinate on the near-horizon region.

Let us proceed as follows: first, let us briefly review Lovelock theory of arbitrary order and in arbitrary dimension, and discuss symmetries near the event horizon in that context. Let us calculate the conserved charges corresponding to the asymptotic vectors that preserve the boundary conditions at the horizon and, at the same time, generate the BMS-like supertranslations there. Then, let us calculate the algebra that these charges satisfy. In particular, let us explicitly check what would be a natural intuition, namely, that higher orders in the curvature do not introduce central terms that were not present in the case of Einstein gravity. Then, let us reproduce the same analysis for the case of extremal horizons, which must be considered separately. The example of the Jackiw-Teitelboim theory will serve to illustrate the form that the charges take at the horizon.

\section{Horizon symmetries}

We will consider the Lovelock theory of gravity, which is the natural extension of Einstein gravity to arbitrary dimension $D$, and is usually considered as a solvable model to investigate the effects of higher-curvature terms. The theory is defined by the action functional
\begin{equation}
    S = \int d^D x \sqrt{-g}\, \sum_{p=1} \frac{\alpha_p}{2^p} \,\,\delta^{\mu_1 \nu_1 \cdots \mu_p \nu_p}_{\rho_1 \sigma_1 \cdots \rho_p \sigma_p} 
    \, \prod_{r=1}^{p}\, R^{\rho_r \sigma_r}_{\ \ \ \mu_r \nu_r}
    ,
\end{equation}
where $\alpha_p$ are coupling constants of mass-dimension $2p-D$ and $\delta^{\mu ...\nu}_{\rho ... \sigma  } $ is the generalized antisymmetric Kronecker delta. We set the convention $\alpha_1=(16\pi G)^{-1}$. The action includes higher-curvature terms up to order $D/2$; the sum over $p$ turns out to be finite, as the terms of order $\mathcal{O}(R^n)$ with $n>D/2$ identically vanish.

We are interested in studying the near horizon form of the metric in $D$ spacetime dimensions. In order to do so, we consider the line element $ds^2=g_{\mu\nu}dx^{\mu}dx^{\nu}$ (with $\mu ,\nu =0,1,2,...,D-1$) written in the following coordinates
\begin{equation}
ds^2=-2\kappa \rho \, dv^2 + 2\, dvd\rho + 2\rho h_i(\rho ,x)\,dx^idv + \hat{g }_{ij}(\rho , x)\,dx^idx^j \label{d2}
\end{equation}
where $x^0=v$ is the null coordinate at the isolated horizon, $x^{D-1}=\rho $ is the distance from the horizon, and $x^i$ (with $i=1,2,...,D-2$) are the spatial coordinates on the spacelike constant-$v$ sections of the horizon. The horizon, $H$, is located at $\rho=0$. The constant-$v$ section of $H$, which will be denoted $\mathcal{H}$, has induced metric $\hat{g}_{ij}$. The constant $\kappa $ is the surface gravity of $H$. Metric components $g_{0i}=\rho h_i$ and $g_{ij}=\hat{g}_{ij}$ are assumed to obey the near-horizon expansion
\begin{eqnarray}
h_i(\rho ,x) &=& h_i^{(1)}(x)+\rho \,h_i^{(2)}(x) +\rho^2\, h_i^{(3)}(x) + ... \\ 
\hat{g }_{ij}(\rho ,x) &=& \hat{g }_{ij}^{(0)}(x)+\rho \, \hat{g }_{ij}^{(1)}(x) +\rho^2\, \hat{g }_{ij}^{(2)}(x) + ... \label{d4}
\end{eqnarray}
where the ellipsis stand for subleading terms, i.e., higher order terms in the $\mathcal{O}(\rho^n)$ expansion. Functions $h_{i}^{(n)}$ and $\hat{g}^{(n)}_{ij}$ depend only on $x^i$.

We consider the asymptotic Killing vector field in the near-horizon region of a black hole in $D$-dimensional spacetime \cite{Donnay}
\begin{equation}
    \chi = \Phi(x)\, \partial_v - \rho \hat{g}^{(0)i}_{\, j}\partial^j\Phi(x)\partial_i+
    \frac{\rho^2}{2}\hat{g}^{(0)}_{ij}h^i\partial^j\Phi(x)\partial_{\rho }+ \mathcal{O}(\rho^3) \label{Chi}
\end{equation}
where $\Phi(x)$ is an arbitrary function of the horizon spatial coordinates $x^i$, and $\hat{g } = \det(\hat{g }^{(0)}_{ij})$ is the determinant of the induced metric on the spatial section of the horizon at $\rho = 0$. Vector $\chi $ leaves the form of the metric (\ref{d2})-(\ref{d4}) invariant, i.e., probably varying the specific functions $h_i^{(n)}$, $\hat{g}^{(n)}_{ij}$ but respecting the expansion in powers of $\rho$ (see the scheme in Figure 1).

Under the supertranslation generated by $\chi$ the metric functions transform as follows
\begin{equation}
\delta_{\chi} h_i = -2\kappa \, \partial _i \Phi \label{Che}
\end{equation}
with $\delta_{\chi}\kappa =0$, $\delta_{\chi}\hat{g }_{ij}=0$. Under diffeomorphisms on $\mathcal{H}$, which are generated by Killing vectors $\xi = Y^{i}(x)\partial_i$, the metric functions transform as follows 
\begin{equation}
\delta _{\xi}\hat{g }_{ij} = \mathcal{L}_{\xi} \hat{g }_{ij} \, , \ \ \ \delta_{\xi}h_i=\mathcal{L}_{\xi}h_i-2\kappa \, \partial_i\Phi \, .
\end{equation}
It is important to note a key distinction from previous studies on horizon symmetries—such as \cite{Donnay, Donnay2}, namely, that we are not restricted to configurations in which rotational symmetries around the horizon are local conformal transformations. Instead, we consider the full group of diffeomorphisms on $\mathcal{H}$. The combination of transformations (\ref{Chi}) and (\ref{Che}) closes an algebra that we will later discuss. 

Before looking at the symmetry algebra, we aim to compute the Noether charge $Q[\chi]$ associated to this vector field $\chi$ evaluated on the spacelike sections of the horizon $\mathcal{H}$ at $\rho = 0$. This can be done by resorting to the phase space formalism \cite{BB}, cf. \cite{IW}. The result for Lovelock gravity reads
\begin{eqnarray}
    Q[\chi] &=& 2\kappa \, \int_{\mathcal{H}}  d^{D-2}x\, \sqrt{\hat{g }}\,  \Phi(x) \sum_{p=1}\Big(  \frac{p\, \alpha_p}{2^{p-1}}\, \delta^{i_1 j_1 ... i_{p-1}j_{p-1}}_{k_1 \ell_1 ... k_{p-1}\ell_{p-1}}\, \nonumber \\
    &&\ \ \ \ \ \ \ \ 
    %\prod_{t=1}^{p-1} \hat{R}^{k_t\ell_t}_{\ \ \ i_t j_t}\Big).
    \hat{R}^{k_1\ell_1}_{\ \ \ i_1 j_1} \hat{R}^{k_2\ell_2}_{\ \ \ i_2 j_2}   ...\, \, \hat{R}^{k_{p-1}\ell_{p-1}}_{\ \ \ i_{p-1} j_{p-1}}\Big)\label{l8}
\end{eqnarray}
where $\hat{R}^{k\ell}_{\ \ i j}$ is the Riemann tensor on $\mathcal{H}$ constructed with the $(D-2)$-dimensional metric $\hat{g }_{ij}^{(0)}$, with $\hat{g }=\det \hat{g }_{ij}^{(0)}$. 

In the particular case $\Phi =1$, the charge (\ref{l8}) reduces to
\begin{eqnarray}
Q[\partial_v ]= \frac{\kappa }{2\pi }\times \frac{A}{4G}+ 4\kappa \alpha_2 \int_{\mathcal{H}}  d^{D-2}x\, \sqrt{\hat{g }} \, \hat{R}\, + ... 
\end{eqnarray}
which coincides with the Wald entropy formula for Lovelock theory \cite{Jacobson}; $A$ being the area of the spacelike section of the horizon. Applied to the static, spherically symmetric black hole solution \cite{BD} in $D=5$, for example, this yields
\begin{eqnarray}
Q[\partial_v ]= \frac{\kappa }{2\pi }\times \frac{2\pi^2r_+(r_+^2+192\pi G \alpha_2)}{4G}.
\end{eqnarray}
with $r_+$ being the horizon radius in spherical coordinates $r=\rho +r_+$. This reproduces the known result for the black hole entropy in Einstein-Gauss-Bonnet theory.
\begin{figure}[ht]
\centering
\includegraphics[scale=0.38]{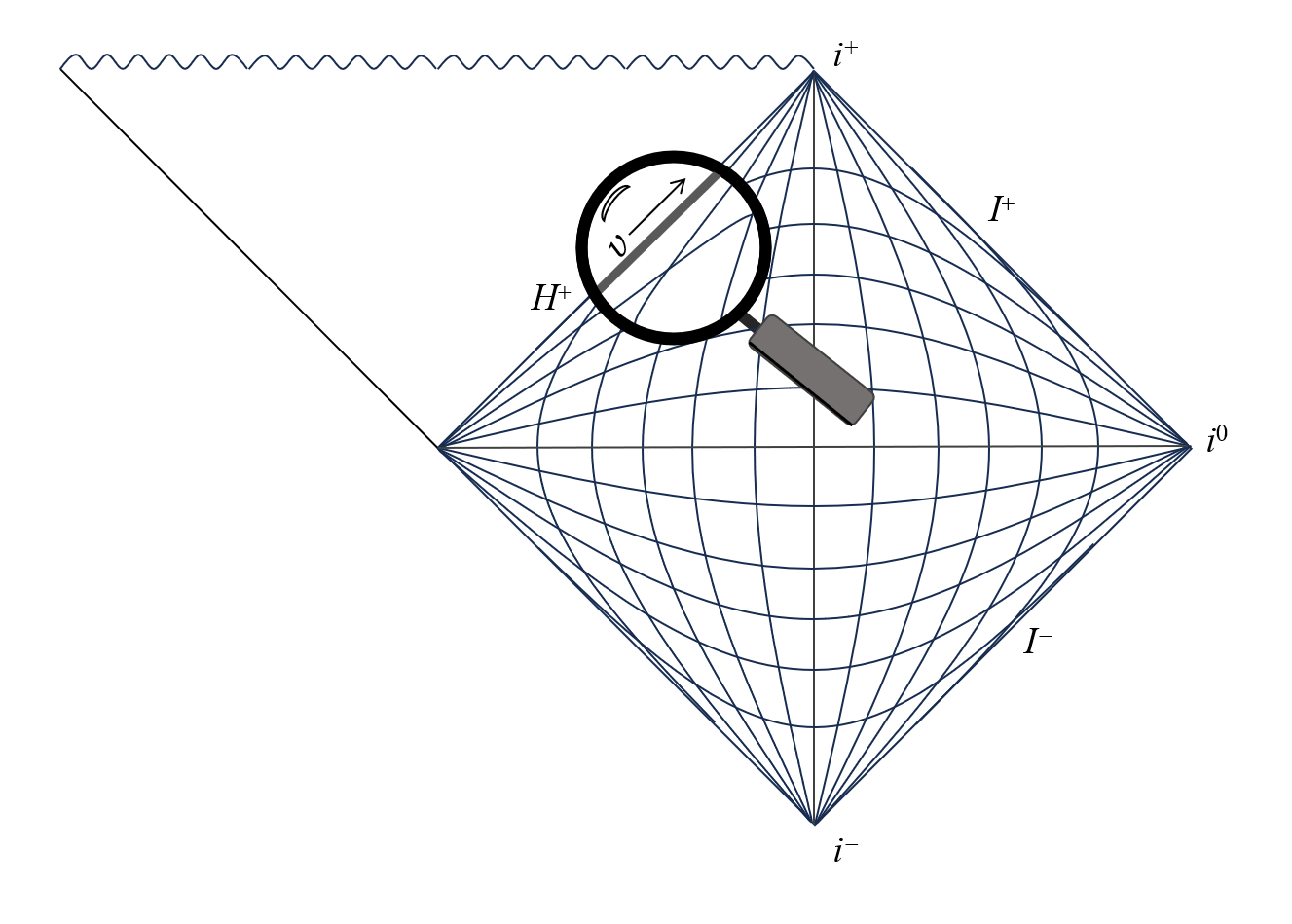}
\caption{Asymptotic Killing vector $\chi =\Phi(x)\partial_v$ generates supertranslations along the null direction $v$ on the horizon, with $x^i$ the coordinates on the constant-$v$ slices of the horizon.} \label{Fig}
\end{figure}

\section{Symmetry algebra}

Now, let us go back to the symmetry algebra. Take the asymptotic Killing vector $\chi  = \Phi(x) \, \partial_v + ...$, which generates the horizon supertranslations, together with the vector fields that generate diffeomorphisms on the horizon, namely $ \xi = Y^i(x)\, \partial_i$.

The algebra of diffeomorphisms is obtained by computing the Lie bracket of the corresponding Killing vectors. For the asymptotic symmetries we are dealing with, this algebra turns out to be infinite-dimensional, with the supertranslations algebra being an Abelian ideal in semi-direct sum with the algebra of diffeomorphisms $\text{Diff}(\mathcal{H})$.

Interesting information is encoded in the algebra of charges, which is obtained by computing the variations $\delta_{\chi_1}Q[\chi_2]$. Schematically, this algebra takes the form \cite{BB}
\begin{equation}
    \{ Q[\chi_1], Q[\chi_2] \} = Q[[\chi_1, \chi_2]] + K(\chi_1, \chi_2),
\end{equation}
with $ K(\chi_1, \chi_2) $ being a possible central extension. In this case, it can be shown explicitly that 
\begin{equation}
K = 0.
\end{equation}

Defining $T[\Phi_a]\equiv Q[\chi _a]$ and $J[Y_a]\equiv Q[\xi _a]$, the algebra of charges can be summarized by the bracket
\begin{equation}
\begin{aligned}
 \{ T[\Phi_1], T[\Phi_2] \} = 0,
\end{aligned}
\end{equation}
together with the brackets
\begin{equation}
\begin{aligned}
\{ J[Y_1], J[Y_2] \} = J[[Y_1, Y_2]], \ \  \{ J[Y_1], T[\Phi_2] \} = T[\mathcal{L}_{Y_1}\Phi_2 ].
\end{aligned}
\end{equation}
This is the semi-direct sum algebra
\[
\text{Diff}(\mathcal{H}) \ltimes C^{\infty }(\mathcal{H}),
\]
where supertranslations form an Abelian ideal. A similar algebra of charges is obtained for the case of extremal horizons, although that case has to be treated separately.

\section{Extremal horizons}

The asymptotic Killing vector that generates the infinite-dimensional near horizon symmetries of the extremal ($\kappa =0$) horizon is given by
\begin{equation}
{\chi }=\Phi(x)\, v\partial_v \, + \, \mathcal{O}(\rho ^2)\, ,
\end{equation}
where, again, $\Phi (x)$ is an arbitrary function of $x^i$. In this case the leading term in $g_{vv}$ is of order $\mathcal{O}(\rho^2)$. Notice that, now, the factor $\chi $ generates a local dilation in $v$.
The corresponding charge is
\begin{eqnarray}
    Q[{\chi} ] &=&  \, \int_{\mathcal{H}}  d^{D-2}x\, \sqrt{\hat{g }}\,  \Phi(x) \sum_{p=1}\Big(  \frac{p\, \alpha_p}{2^{p-2}}\, \delta^{i_1 j_1 ... i_{p-1}j_{p-1}}_{k_1 \ell_1 ... k_{p-1}\ell_{p-1}}\, \nonumber \\
    &&\ \ \ \ \ \ 
    %\prod_{t=1}^{p-1} \hat{R}^{k_t\ell_t}_{\ \ \ i_t j_t}\Big).
    \hat{R}^{k_1\ell_1}_{\ \ \ i_1 j_1} \hat{R}^{k_2\ell_2}_{\ \ \ i_2 j_2}   ...\, \, \hat{R}^{k_{p-1}\ell_{p-1}}_{\ \ \ i_{p-1} j_{p-1}}\Big)\label{carguita}
\end{eqnarray}
This means that, while for $\kappa =0$ charge (\ref{l8}) identically vanishes, extremal horizons can still have non-trivial charges associated to horizon superdilations in $v$, cf. \cite{Donnay2}. 

In the particular case $\Phi =1$, we obtain
\begin{eqnarray}
    Q[v\partial_v ] &=&  \, \frac{\hbar }{2\pi }\, S\, +\,   ...
\end{eqnarray}
where $S$ is the Bekenstein-Hawking entropy, and the ellipsis stand for higher-curvature corrections to it. For example, in $D=5$ we can take the quadratic theory with positive cosmological constant $\Lambda $, and then consider the Nariai limit of the asymptotically de Sitter (dS) static black hole solution. In that limit, the horizon radius $r_+$ is given by
\begin{equation}
r_+=\sqrt{\frac{3}{\Lambda}}=\frac{L^2}{\sqrt{2L^2+64\pi G \alpha_2}}\, ,
\end{equation}
with $L$ being the effective curvature radius of the spacetime. This yields the charge
\begin{equation}
Q[v\partial_v]=\frac{\pi}{4G} \sqrt{\frac{27}{\Lambda^3}}+48\pi^2\alpha_2\sqrt{\frac{3}{\Lambda}},
\end{equation}
which correctly reproduces the entropy of the Nariai horizon in $D=5$ with finite $\alpha_2$ corrections. The latter correction vanishes for locally dS$_5$ solutions with spatially flat $3$-dimensional sections. The case of extremal charged black hole solutions in $D$ dimensions can also be worked out explicitly in a similar way. 

\section{Remarks}

The conserved charges associated to the near horizon supertranslation symmetry can be computed in gravitational theories of arbitrary spacetime dimension and involving curvature terms of any order. In Lovelock theory these charges take the form (\ref{l8}) and (\ref{carguita}), i.e., the form of nested Lagrangian densities corresponding to topological invariants, each weighted by the supertranslation function -- thus providing a natural generalization of the Wald entropy formula. 

In $D=4$ dimensions, provided one keeps the quadratic term in the action, the computation of the supertranslation charge reduces to the evaluation of the JT gravity action on the constant-$v$ slices of the horizon; namely
\begin{equation}
Q[\chi ] = \frac{1 }{16\pi G} \int_{\mathcal{H}}d^2x \sqrt{\hat{g}} \, \Phi(x)\, \Big( \hat{R} -2\Lambda  \Big) 
\end{equation}
where the supertranslation function $\Phi(x)$ plays the role of the JT field, the two-dimensional Newton constant and the cosmological constant are given by $G = 1 /(64\pi \alpha_2 )$ and $\Lambda = -\alpha_1/(4\alpha_2)$.

Whether the horizon charge takes the form of the JT action depends on the structure of the 4-dimensional gravitational theory, including the quadratic Lanczos term. In 4 dimensions, this term corresponds to the Euler characteristic and arises naturally within the Lovelock hierarchy. The same feature extends to higher (even) dimensions. The JT case thus serves as a useful example to illustrate the general structure that horizon charges can exhibit. Given the renewed interest in JT gravity as a toy model for exploring aspects of quantum gravity, black holes, and holography, it is hard not to speculate whether the emergence of the JT action as the supertranslation charge at the horizon hints at a deeper physical interpretation in the holographic context. While such an interpretation remains unclear, we believe the observation is worth highlighting. It is also worth emphasizing that the same form of the horizon charge arises for both non-extremal and extremal horizons, provided the appropriate asymptotic Killing vector is identified.

\[\]

\subsection*{Acknowledgments}
This work is the extension to higher-curvature gravity of the results reported in references \cite{Donnay, Donnay2}; G.G. thanks Laura Donnay, Hern\'an Gonz\'alez and Miguel Pino for previous collaboration in the subject. The work of M.C. was supported by DGAPA UNAM grant IN116823. The work of J.O. was supported by FONDECYT grant 1221504.

\[ \]

%%%%%%%%%%%%%%%%%%%%%%%%%%%% BIBLIOGRAPHY %%%%%%%%%%%%%%%%%%%%%%%%%%%%%%%%%%%%%%
\providecommand{\href}[2]{#2}\begingroup\raggedright\endgroup
\end{document}